\documentclass[10pt,journal]{IEEEtran}
\usepackage{amsrefs}
\newenvironment{rezabib}
  {\bibdiv\biblist\setupbib}
  {\endbiblist\endbibdiv}

\def\setupbib{\catcode`@=\active}
\begingroup\lccode`~=`@
  \lowercase{\endgroup\def~}#1#{\gatherkey{#1}}
\def\gatherkey#1#2{\gatherkeyaux{#1}#2\gatherkeyaux}
\def\gatherkeyaux#1#2,#3\gatherkeyaux{\bib{#2}{#1}{#3}}

\usepackage{ulem}
\usepackage{times}
\usepackage{epsfig}
\usepackage{graphicx}
\usepackage{amsmath}
\usepackage{amssymb}
\usepackage{amsmath}
\DeclareMathOperator*{\argmin}{argmin}
\begin{document}

\title{Coherent Point Drift Networks: Unsupervised Learning of Non-Rigid Point Set Registration}

\author{Lingjing~Wang,
        Xiang~Li,
        Jianchun~Chen,
        and~Yi~Fang
\IEEEcompsocitemizethanks{L.Wang is with MMVC Lab, the Department of Mathematics, New York University, New York, NY, 30332 USA, e-mail: lingjing.wang@courant.nyu.edu. X.Li is with the MMVC Lab, New York University, New York, NY11201, USA, e-mail: lixiang709709@gmail.com. J.Chen is with the MMVC Lab, New York University, New York, NY11201, USA, e-mail: jc7009@nyu.edu. Y.Fang is with MMVC Lab, Dept. of ECE, NYU Abu Dhabi, UAE and Dept. of ECE, NYU Tandon School of Engineering, USA, e-mail: yfang@nyu.edu.}
\IEEEcompsocitemizethanks{Corresponding author: Yi Fang. Email: yfang@nyu.edu}}
\maketitle

\begin{abstract}
Given new pairs of source and target point sets, standard point set registration methods often repeatedly conduct the independent iterative search of desired geometric transformation to align the source point set with the target one. This limits their use in applications to handle the real-time point set registration with large volume dataset. This paper presents a novel method, named coherent point drift networks (CPD-Net), for the unsupervised learning of geometric transformation towards real-time non-rigid point set registration. In contrast to previous efforts (e.g. coherent point drift), CPD-Net can learn displacement field function to estimate geometric transformation from a training dataset, consequently, to predict the desired geometric transformation for the alignment of previously unseen pairs without any additional iterative optimization process. Furthermore, CPD-Net leverages the power of deep neural networks to fit an arbitrary function, that adaptively accommodates different levels of complexity of the desired geometric transformation. Particularly, CPD-Net is proved with a theoretical guarantee to learn a continuous displacement vector function that could further avoid imposing additional parametric smoothness constraint as in previous works. Our experiments verify the impressive performance of CPD-Net for non-rigid point set registration on various 2D/3D datasets, even in the presence of significant displacement noise, outliers, and missing points. Our code will be available at https://github.com/nyummvc/CPD-Net.
\end{abstract}





%


\IEEEPARstart{P}{oint} set registration is a fundamental computer vision task, which has wide applications in many fields such as image registration, object correspondence, large-scale 3D reconstruction and so on \cite{bai2007skeleton,bai2008path,myronenko2009image,ma2016non,wu2012online,klaus2006segment,maintz1998survey,besl1992method,raguram2008comparative,yuille1988computational,sonka2014image}. Existing approaches often solve the registration problem through an iterative optimization process to search the optimal geometric transformation to minimize a pre-defined alignment loss between transformed source point set and target point set\cite{myronenko2007non,jian2011robust,ma2013robust,ma2014robust,ling2005deformation}. The geometric transformation can be modeled by a specific type of transformation (e.g. thin plate spline) \cite{besl1992method}, or described as a model-free deformation field \cite{myronenko2007non} where a displacement vector function is defined. The rotation, translation and scaling are often used to model rigid geometric transformation, while the affine and thin plate spline are widely used to model non-rigid geometric transformation. For model-based methods, the registration task turns into a searching process for an optimal set of parameters of the model. For model-free method, the registration task turns into to process to determine the displacement field that moves the source point set towards the target one. Regardless of the model-based or model-free point set registration approaches, most existing efforts have to start over a new iterative optimization process for each pair of input to determine the geometric transformation. Consequently, the intensive computation in the iterative routine 
\begin{figure}[h!]
\centering
\includegraphics[width=9cm]{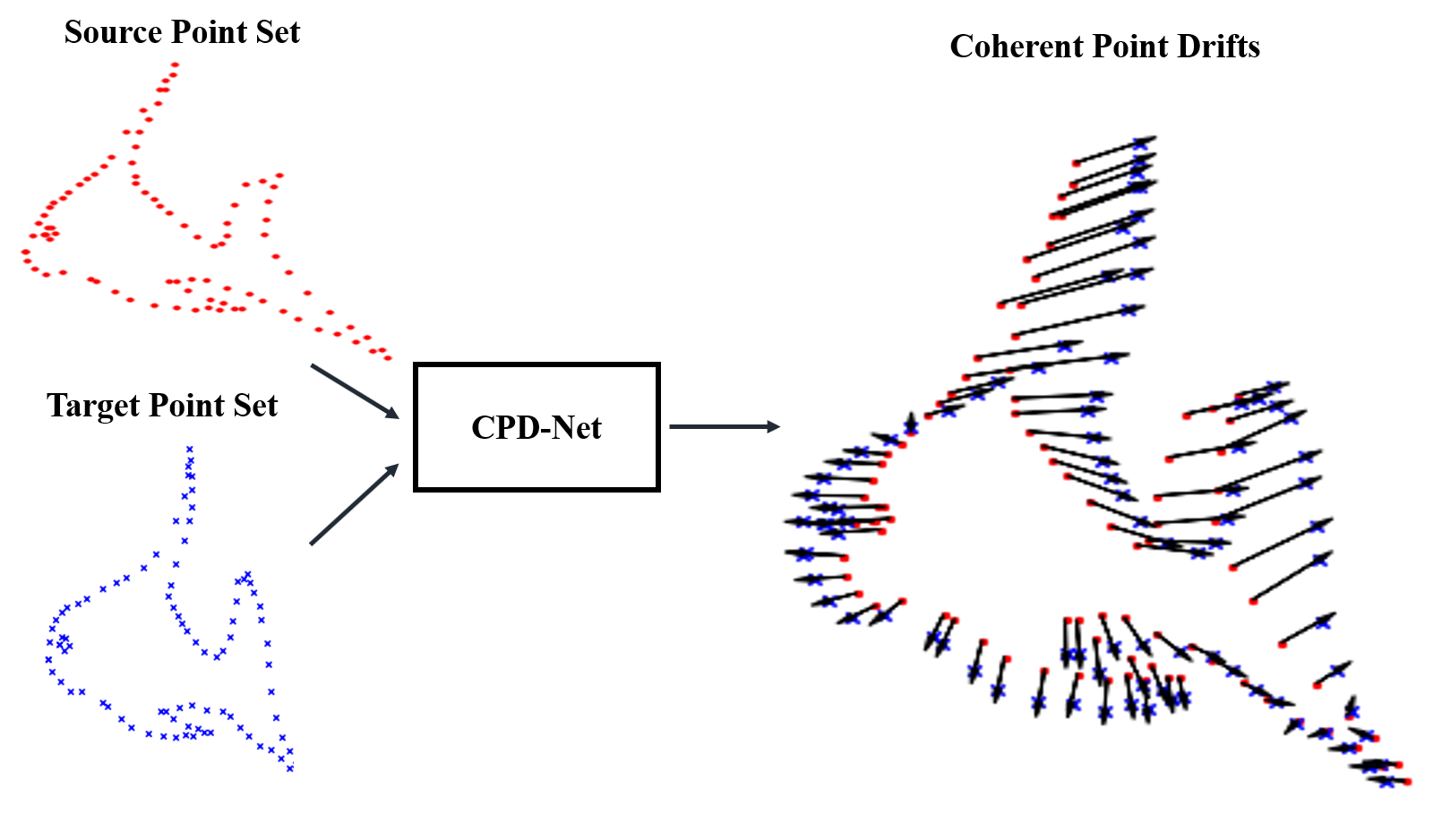}
\caption{For a given pair of source and target point sets, CPD-Net directly predicts the coherent drifts to move source point set towards target one.}
\label{c1}
\end{figure}
pose great challenges for existing approaches to handle a large scale dataset in real-time application. To address this challenge, we are motivated to develop a learning based approach for point set registration and to generalize its ability from training process to predict the desired geometric transformation to register previously unseen pair of point sets.

\begin{figure*}
\centering
\includegraphics[width=17.5cm,height=7.5cm]{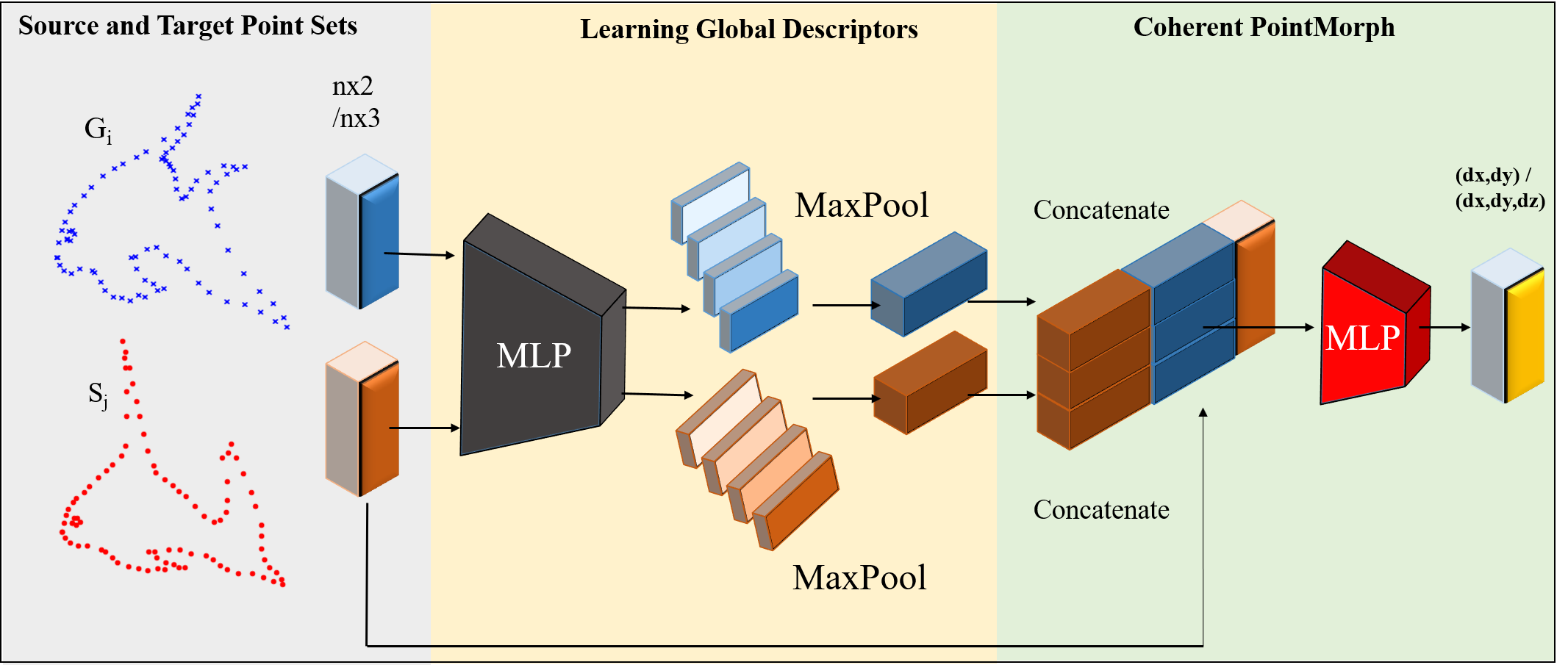}
\caption{Our pipeline. The proposed structure includes three parts: learning shape descriptor, coherent PointMorph, and the alignment Loss. For a pair of source point set $\bold{S_i}$ and target point set $\bold{G_j}$, we firstly leverage MLPs to learn two global descriptors $\bold{L_{S_i}}$ and $\bold{L_{G_j}}$. We then concatenate these two descriptors to each coordinate $\{x_k\}_{k=1,2,...,n}$ of source points as the input ($[\bold{x_k},\bold{L_{S_i}},\bold{L_{G_j}}]$) for PointMorph structure. We further use MLPs to learn the drifts for each source point. Finally, we move the source point set by our predicted drifts and define the alignment loss function between target and transformed source point sets for back-propagation.}
\label{fig222}
\end{figure*}
In this paper, we proposed a novel approach, named coherent point drift networks (CPD-Net), for the unsupervised learning of displacement vector function to estimate a model-free geometric transformation for point set registration. Figure \ref{fig222} illustrates the pipeline of the proposed CPD-Net which consists of three major components. The first component is ``Learning Shape Descriptor'. In this component, the global shape descriptor is learned with a multilayer perceptron (MLP). The second component is ``Coherent PointMorph''. In this component, we firstly concatenate the global shape descriptor of the source point set and target point set with the coordinate of source point to form a geometric transformation descriptor for the input pair. Three successive MLP are then defined and take the geometric transformation descriptor as input to learn the continuous displacement vector field. The third component is ``Point Set Alignment''. In this component, a loss function is defined to assess the quality of alignment. The main contributions of our proposed method as follows:

\begin{itemize}

\item This paper introduces a novel Coherent Point Drift Networks (CPD-Net) for the unsupervised learning of point set registration. The proposed network can be generalized to predict geometric transformation for non-rigid point set registration.\\

\item CPD-Net leverages the power of deep neural networks to fit an arbitrary function, thus makes it possible to accommodate different levels of complexity of the target geometric transformation for best aligning the pair of point sets.\\

\item The CPD-Net theoretically guarantees the continuity of predicted displacement field as geometric transformation, which naturally eliminates the necessity to impose a parametric hand-crafted smoothness constraint.\\

\item The CPD-Net is free of specific geometric model selection for modeling the desired transformation, which avoids the potential mismatch between the transformation described by specific pre-defined geometric models and the actual transformation required for point set registration.

\end{itemize}

\section{Related Works}
\subsection{Iterative registration methods.} 
The development of optimization algorithms to estimate the rigid or non-rigid geometric transformations in an iterative routine have been attracting great research attention in decades.
The Iterative Closest Point (ICP) algorithm \cite{besl1992method} is one successful solution for rigid registration. It initializes an estimation of a rigid function and then iteratively chooses corresponding points to refine the transformation.  However, the ICP algorithm is reported to be vulnerable to the selection of corresponding points for initial transformation estimation, and also incapable of dealing with non-rigid transformation. To accommodate the deformation (e.g. morphing, articulation) between a pair of point sets, many efforts were spent on the development of algorithms for a non-rigid transformation. For non-rigid registration, classical previous methods can usually be divided into parametric and non-parametric by the target transformation. As a robust parametric method, TPS-RSM algorithm was proposed by Chui and Rangarajan \cite{chui2000new} to estimate parameters of non-rigid transformation with a penalization on second-order derivatives. As a classical non-parametric method, coherence point drift (CPD) was proposed by Myronenko et al. \cite{myronenko2007non}, which successfully introduced a process of fitting Gaussian mixture likelihood to align the source point set with the target one. With the penalization term on the velocity field, the algorithm enforces the motion of source point set to be coherent during the registration process. Another non-parametric vector field consensus algorithm was proposed by Ma et al. \cite{ma2014robust}. This algorithm is emphasized to be robust to outliers. Ma et al. proposed the importance to leverage both local and global structures during the non-rigid point set registration in \cite{ma2016non}. The existing algorithms achieved great success for the registration task. Even though all these methods state the registration task as an independent optimization process for every single pair of source and target point sets, they greatly inspire us for designing our learning-based system.

\subsection{Learning-based registration methods.} In recent years, learning-based methods achieved great success in many fields of computer vision \cite{su2015multi,sharma2016vconv,maturana2015voxnet,bai2016gift,qi2017pointnet,verma2018feastnet,masci2015geodesic,zeng20173dmatch}. Especially recent works started a trend of directly learning geometric features from cloud points (especially 3D points), which motivates us to approach the point set registration problem using deep neural networks by leveraging the possibility of direct learning features from points \cite{rocco2017convolutional,balakrishnan2018unsupervised,zeng20173dmatch,qi2017pointnet,verma2018feastnet,masci2015geodesic}. The closest research to this work is the unsupervised non-parametric non-rigid registration of 3D volumetric medical image proposed by Balakrishnan et al. \cite{balakrishnan2018unsupervised}. In this research, authors use a voxelMorph CNN architecture to learn the registration field to align two volumetric medical images. Even though, both of our researches engage in learning a non-parametric transformation. In contrast, we discuss a broader task for 2D/3D point set registrations in this paper. In comparison to volumetric data, point clouds have very different characteristics and we target to predict the coherent drifts. Moreover, Our proposed model does not require additional penalization term on the smoothness of predicted transformation. For registration of 2D images, an outstanding registration model was proposed by Rocco et al. \cite{rocco2017convolutional}. This work mainly focuses on parametric approach for 2D image registration. The parameters of both rigid and non-rigid function (TPS) can be predicted by a CNN-based structure from learning the correlation relationship between a pair of source and target 2D images. Even though these learning-based methods are not exactly designed for solving the point set registration problem, they are closely related to our paper and greatly encourage us to investigate the efficient learning-based structure for model-free point set registration. 

\section{Approach}
We introduce our approach in the following sections. Firstly we define the learning-based registration problem in section 3.1. In section 3.2, we introduced the module for learning shape descriptor from 2D/3D source/target point sets. Section 3.3 illustrates coherent PointMorph module for learning the smooth drifts to align the source point set with the target one. In section 3.4, The definition of the loss function is provided. The model configurations and the settings for training are described in section 3.5. 

\subsection{Problem statement}
For iterative methods, the optimization task is defined on a given pair of source and target point sets $(\bold{S_i}, \bold{G_j})$. Here, unlike interactive methods, for a given training dataset $\bold{D}=\{(\bold{S_i}, \bold{G_j}) \text{ ,where } \bold{S_i}, \bold{G_j} \subset \mathbb{R}^N (N=2 \text{ or } N=3) \}$, we need to redefine the new optimization task. We assume the existence of a function $g_{\theta}(\bold{S_i},\bold{G_j}) = \phi$ using a neural network structure, where $\phi$ is the coherent drifts for each (sampled) point in source point set towards its corresponding point in the target point set. In Figure \ref{fig222}, we illustrate the overview of our proposed network structure. Taking a pair of source and target point sets $(\bold{S_i},\bold{G_j})$ as input, out model compute the coherent point drifts $\phi$ based on a set of parameters $\theta$, which are weights in the neural network structure. We further transform the input source points by simply moving the drifts and then, measure the similarity between the transformed source and target point sets as the alignment loss to update $\theta$.

For a given dataset, we use stochastic gradient descent based algorithm to optimize parameters set $\theta$ for minimizing the expected loss function:
\begin{equation}
\begin{split}
\bold{\theta^{optimal}} =\argmin_{\theta}[\mathbb{E}_{(\bold{S_i},\bold{G_j})\sim \bold{D}}[\mathcal{L}(\bold{S_i},\bold{G_j}, g_{\theta}(\bold{S_i},\bold{G_j}))]],
\end{split}
\end{equation}

, where $\mathcal{L}$ represents a similarity measure. CPD-Net does not require any ground truth drifts for supervision. For an unseen pair from the testing dataset, our model directly predicts the desired coherent drifts without any additional iterative process.  

\subsection{Learning shape descriptor}

For a given input point set, we firstly learn a shape descriptor that captures representative and deformation-insensitive geometric features. Let $(\bold{S_i},\bold{G_j})$ denotes the input source and target point sets and $(\bold{L_{S_i}},\bold{L_{G_j}})$ denotes their shape descriptor, where $\bold{L_{S_i}}, \bold{L_{G_j}} \subset \mathbb{R}^m $ as shown in Figure \ref{c1}. To address the problem of irregular format of point set, we introduce the following encoding network, which includes $t$ successive multi-layer perceptrons (MLP) with ReLu activation function $\{f_i\}_{i=1,2,...,t}$, such that: $f_i : \mathbb{R}^{w_{i}}\to \mathbb{R}^{w_{i+1}}$, where $w_{i}$ and $w_{i+1}$ are the dimension of input layer and output layer. The encoder network is defined as: $\forall (\bold{S_i},\bold{G_j})$, 
\begin{equation}
\begin{split}
\bold{L_{S_i}}= \text{Maxpool} \{ f_tf_{t-1}...f_1(\bold{x_i})\}_{\bold{x_i}\in \bold{S_i}}
\end{split}
\end{equation}
\begin{equation}
\begin{split}
\bold{L_{G_j}}= \text{Maxpool} \{ f_tf_{t-1}...f_1(\bold{x_i})\}_{\bold{x_i}\in \bold{G_j}}
\end{split}
\end{equation}

We use the Maxpool function to extract the order-invariant descriptors for both the source and the target point sets. The readers can refer to PointNet \cite{qi2017pointnet} for detailed discussion. One can also use other symmetric operators such as summation, average pooling function and so on. This structure can be easily adapted for 3D point set inputs. Other point signature learning architecture such as PointNet++ \cite{qi2017pointnet++} can be easily implemented here as well.
\begin{figure}[h!]
\centering
\includegraphics[width=8cm]{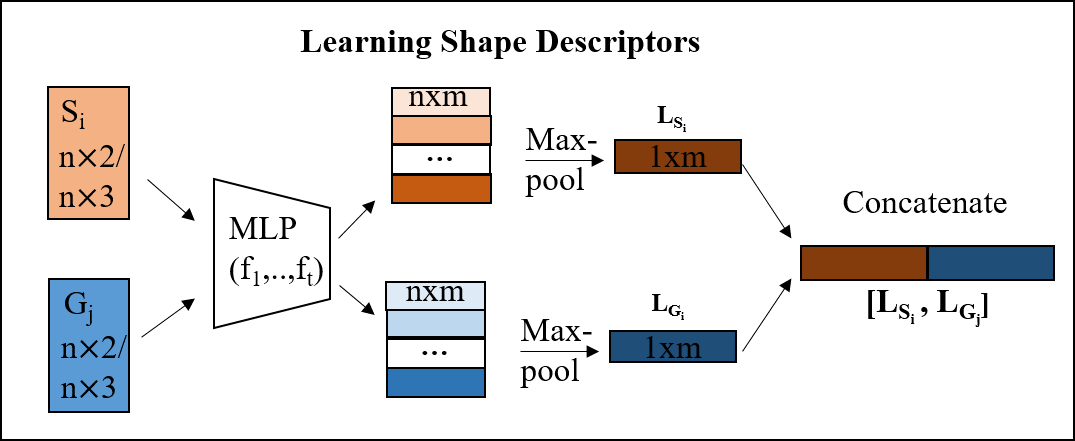}
\caption{Schema of learning shape descriptor tensor process.}
\label{c1}
\end{figure}

\subsection{Coherent PointMorph architecture}

In general, the geometric transformation $\mathcal{T}$ for each point $x$ in a point set $\bold{X} \subset \mathbb{R}^2/\mathbb{R}^3$ can be defined as: 
\begin{equation}
\label{eq:31}
\begin{split}
\mathcal{T}(x,v) = x + v(x)
\end{split}
\end{equation} 
, where $v: \mathbb{R}^2 \to \mathbb{R}^2 /\mathbb{R}^3 \to \mathbb{R}^3$ is a ``point displacement'' function. The point set registration task requires us to determine this displacement function $v$ such that the source point set can be coherently moved towards target point set. In addition to the desired perfect alignment after registration, it is necessary for function $v$ to be a continuous and smooth function according to the Motion Coherent Theory (MCT) \cite{yuille1989mathematical}. Fortunately, by leveraging deep neural networks, we can construct the suitable displacement function $v$ via our proposed coherent pointMorph architecture, which can not only align the source and target point sets but also satisfies the continuous and smooth characteristics.

\begin{figure}[h!]
\centering
\includegraphics[width=7cm]{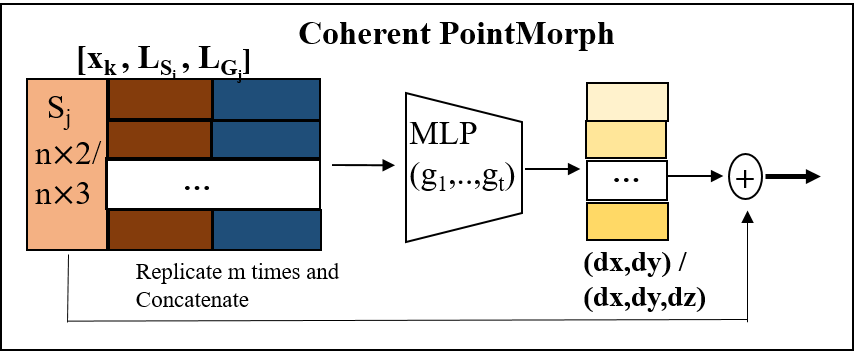}
\caption{The schema of learning coherent PointMorph process.}
\label{c2}
\end{figure}

 As illustrated in Figure \ref{c2}, given $n$ 2-D/3-D points in the source point set, we firstly duplicate the concatenated descriptors learned from source and target point sets $[\bold{L_{S_i}},\bold{L_{G_j}}]$ for $n$ times. Then, the low-dimensional coordinate of each point in the source point set is concatenated with the joint global descriptors as $[\bold{x_i}, \bold{L_{S_i}},\bold{L_{G_j}}]$. Using the joint global descriptors as inputs, we further define a PointMorph MLP (multi-layer perceptrons) architecture for learning the coherent point drifts to move the source point set towards alignment with the target one. This architecture includes successive multi-layer perceptrons (MLP) with ReLu activation function: $\{g_i\}_{i=1,2,...,s}$, such that: $g_i : \mathbb{R}^{v_{i}}\to \mathbb{R}^{v_{i+1}}$, where $v_{i}$ and $v_{i+1}$ are the dimension of input layer and output layer. The desired displacement function is formulated as $v = g_sg_{s-1}...g_1$. Therefore, $\forall (\bold{S_i},\bold{G_j})$,
\begin{equation}
\begin{split}
\bold{dx_i}=g_sg_{s-1}...g_1([\bold{x_i},\bold{L_{S_i}},\bold{L_{G_j}}])
\end{split}
\end{equation}
\begin{equation}
\begin{split}
\bold{S_i}'=\phi(\bold{S_i})= \{\bold{x_i}+ \bold{dx_i}\}_{\bold{x_i}\in \bold{S_i}}
\end{split}
\end{equation}

,where $\bold{S_i'}$ is the transformed source point set and $\bold{dx_i}$ represents the predicted drift for each point $\bold{x_i} \in \bold{S_i}$. The notation [*,*] represents the concatenation of vectors in the same domain. We should notice that due to the high nonlinearity of neural networks, there is no difficulty to minimize the similarity loss function between $\bold{S'}$ and $\bold{G}$. However, the main challenge is to make sure the predicted drifts are coherent \cite{myronenko2007non}. The coherency of drifts is quite important for holding reasonable correspondence and points interpolation for registering large scale point sets as well. Most previous methods deal with this problem by adding a penalization term on smoothness to trade-off target alignment loss for smoothness \cite{balakrishnan2018unsupervised}. Furthermore, we briefly prove the continuity and smoothness of our pointMorph architecture to be used as our deep learning-based solution for the displacement function $v$. We ignore batch normalization layer since it doesn't influence the continuity and smoothness.

\noindent \textbf{Continuity}. 
Since both MLP and activation function are continuous, the continuity of the pointMorph network $\mathcal{F}_{\theta}$ can be trivially proved as a composite of continuous functions. Since $[\bold{L_{S_i}},\bold{L_{G_j}}]$  is concatenated to each point in $\bold{X}$, this concatenation operation does not change the continuity for the pointMorph structure as our displacement function. In contrast, commonly used learning paradigm \cite{balakrishnan2018unsupervised}, which directly maps high dimensional feature space to 2-D displacement field, is unable to guarantee the continuity of displacement field.

\textit{Claim:} For a single layer perceptron with ReLu activation function $g$, $\forall \bold{x_i}, \bold{x_j} \in \bold{S}, \forall \varepsilon >0, \exists \delta>0, $ such that $||\bold{x_i}-\bold{x_j}||< \varepsilon \Longrightarrow ||\bold{dx_i}-\bold{dx_j}||<\delta$, where $\bold{dx_i}$ is defined as $\bold{dx_i}=g([\bold{x_i},\bold{L_{S}},\bold{L_{G}])}$ similarly in equation (6).

\textit{Proof in sketch.} Since $g$ is a linear function with ReLu activation and we assume its weights $\bold{w}$ converges to $\bold{w'}$ after training. $||\bold{x_i}-\bold{x_j}||=||[\bold{x_i},\bold{L_{S}},\bold{L_{G}}]-[\bold{x_j},\bold{L_{S}},\bold{L_{G}}]||$ since $\bold{L_{S}},\bold{L_{G}}$ are identical for each $\bold{x_i}\in \bold{S}$. Since $\bold{w'}$ is constant, $\exists C>0$, s.t.$||\bold{x_i}-\bold{x_j}||>C||\bold{w'}[\bold{x_i},\bold{L_{S}},\bold{L_{G}}]-\bold{w'}[\bold{x_j},\bold{L_{S}},\bold{L_{G}}]||$. Since function $g$ is continuous, if $\bold{w'}[\bold{x_i},\bold{L_{S}},\bold{L_{G}}]\ge 0$, $\exists \delta_1>0$ such that $\forall \bold{x_i}$, if $ ||\bold{x_i}-\bold{x_j}||<\delta_1, \bold{w'}[\bold{x_j},\bold{L_{S}},\bold{L_{G}}]\ge 0$. Similarly if $\bold{w'}[\bold{x_i},\bold{L_{S}},\bold{L_{G}}]\le 0$, $\exists \delta_2>0$ such that $\forall \bold{x_j}$, if $ ||\bold{x_i}-\bold{x_j}||<\delta_2, \bold{w'}[\bold{x_j},\bold{L_{S}},\bold{L_{G}}]\le 0$. Therefore, we pick $\delta=\min(\delta_1, \delta_2, \varepsilon/C)$, $||\bold{dx_i}-\bold{dx_j}||=||g([\bold{x_i},\bold{L_{S}},\bold{L_{G}}])-g([\bold{x_j},\bold{L_{S}},\bold{L_{G}}])||=||\max(\bold{w'}[\bold{x_i},\bold{L_{S}},\bold{L_{G}}], 0)-\max(\bold{w'}[\bold{x_j},\bold{L_{S}},\bold{L_{G}}],0)||=\max(||\bold{w'}[\bold{x_i},\bold{L_{S}},\bold{L_{G}}]-\bold{w'}[\bold{x_j},\bold{L_{S}},\bold{L_{G}}]||, 0)<\max(\frac{1}{C}||\bold{x_i}-\bold{x_j}||,0)<\max(0,\varepsilon/C)=\varepsilon/C < \delta$. \\

\noindent \textbf{Smoothness v.s. Complexity} After choosing a smooth function SoftPlus \cite{nair2010rectified} as the activation function in our pointMorph architecture, it becomes trivial to estimate its smoothness since the displacement function is a composite of a number of smooth functions (MLP and SoftPlus). In practice, Regularization Theory (RT) \cite{girosi1995regularization} uses the oscillatory behavior of a function to further measure the smoothness of displacement function. The desired displacement function is estimated by minimizing Eq.\ref{eq:34} with a balance of the fitting error and the oscillatory measure, which is defined as the norm of displacement function $v$ in Hilbert Space.
\begin{equation}
\label{eq:34}
\begin{split}
E(v) = \sum_{i=1}^{K}{||y_i-v(x_i)||^2}+\lambda \int_{\mathbb{R}}{\sum_{k=0}^{m}{||\frac{\partial^kv}{\partial x^k}||^2}}{dx}
\end{split}
\end{equation}
, where $y_i$ and $x_i$ are corresponding points from target point set $\bold{Y}$ and source point set $\bold{X}$. $\lambda$ is the balancing parameter.

Regarding the ``fitting error'' term in Eq.\ref{eq:34}, our proposed displacement function (pointMorph) $v$ is a multi-layer deep neural structure. According to universal approximation theory, the neural network is able to accommodate arbitrary function by adjusting the number of hidden neurons. Compared with TPS-based function fitting, it is more flexible to control the complexity of the displacement function by hidden neurons than explicitly positioning the TPS controlling points. Regarding the ``oscillatory measure'' term in Eq.\ref{eq:34}, authors in \cite{zhang2016understanding} suggest that with a certain number of weight parameters, the neural network is capable of perfectly fitting any function (e.g. over-fit) by training on a given sample space. In our case, the ``over-fitting'' to high-frequency displacement vector causes the function oscillatory, as result of the introduction of higher norm of $v$. Fortunately, given the popularity of deep learning models, recent research community has been proposing training strategies (e.g. dropout, weight decay) to prevent deep learning models from over-fitting, which naturally help our Displacement Field Predictor network to reduce the risk of predicting an oscillatory displacement function.

\subsection{Loss functions}
In this part, we define the similarity measure between the transformed source point set $\phi(\bold{S_i})$ and the target point set $\bold{G_j}$ as both our loss function and evaluation metric. For two point sets, due to the absence of the corresponding relationship for each point, we cannot adopt the pixel-wise loss in image registration. Fan et al. \cite{fan2017point} first proposed Chamfer Distance (C.D.), which is widely used in practice. We define the Chamfer loss on our transformed source point set $\bold{S}'$ and target points set $\bold{G} $ as:
\begin{equation} 
\begin{split}L(\bold{S'},\bold{G}|\theta)
 &= \sum_{x\in \bold{S'}}\min_{y \in \bold{G}}||x-y||^2_2\\
 &+ \sum_{y\in \bold{G}}\min_{x \in \bold{S'}}||x-y||^2_2
\end{split}
\end{equation}
where $\theta$ represents all the weights in the our network structure. Without specific mention, we use Chamfer loss (C.D.) to validate the performance of our proposed method in this paper. For dataset in presence of outliers and missing points noise, we use the following clipped Chamfer loss:
\begin{equation} 
\begin{split}L(\bold{S'},\bold{G}|\theta)
 &= \sum_{x\in \bold{S'}}\max(\min_{y \in \bold{G}}||x-y||^2_2, c)\\
 &+ \sum_{y\in \bold{G}}\max(\min_{x \in \bold{S'}}||x-y||^2_2,c)
\end{split}
\end{equation}
where c is a hyper-parameter to choose. In our experiment 3 (Section D), we set c equal to 0.1.


\subsection{Settings of CPD-Net}
We train our network using batch data form training data set $\{(\bold{S^i},\bold{G^i}) | (\bold{S^i}, \bold{G^i}) \in \bold{D} \}_{i=1,2,...,b}$, where b denotes the batch size and is set to $16$. As we explain in section 3.2, for learning the shape descriptor tensor, the input is $n\times4/6$ matrix and we use $5$ MLP layers with dimensions $(16,64,128,256,512)$ and a Maxpool layer to convert it to a 512-dimensional descriptor. For learning the coherent PointMorph discussed in 3.3, we use three layers MLP with dimension $(256,128,2/3)$. We use ReLU activation function and implement batch normalization \cite{ioffe2015batch} for every layer except the output layer. Learning rate is set as 0.0001 with 0.995 exponential decay with Adam optimizer. The model is trained on single Tesla K80 GPU.

\section{Experiments}
In this section, we carry out a set of experiments for non-rigid point set registration and assess the performance of our proposed CPD-Net. In section 4.1, we describe the details of datasets that are used for training and testing of CPD-Net. We report the experimental results to evaluate the performance of our trained CPD-Net on 2D and 3D datasets in section 4.2 and 4.3. Section 4.4 discusses the resistance of CPD-Net to various types of noise. In section 4.5, we compare CPD-Net with non-learning based method. In section 4.6, we demonstrate that our model can be used for real-world 3D shape registration. 

\subsection{Experimental Dataset}
\begin{figure*}
\centering
\includegraphics[width=12cm,height=9.5cm]{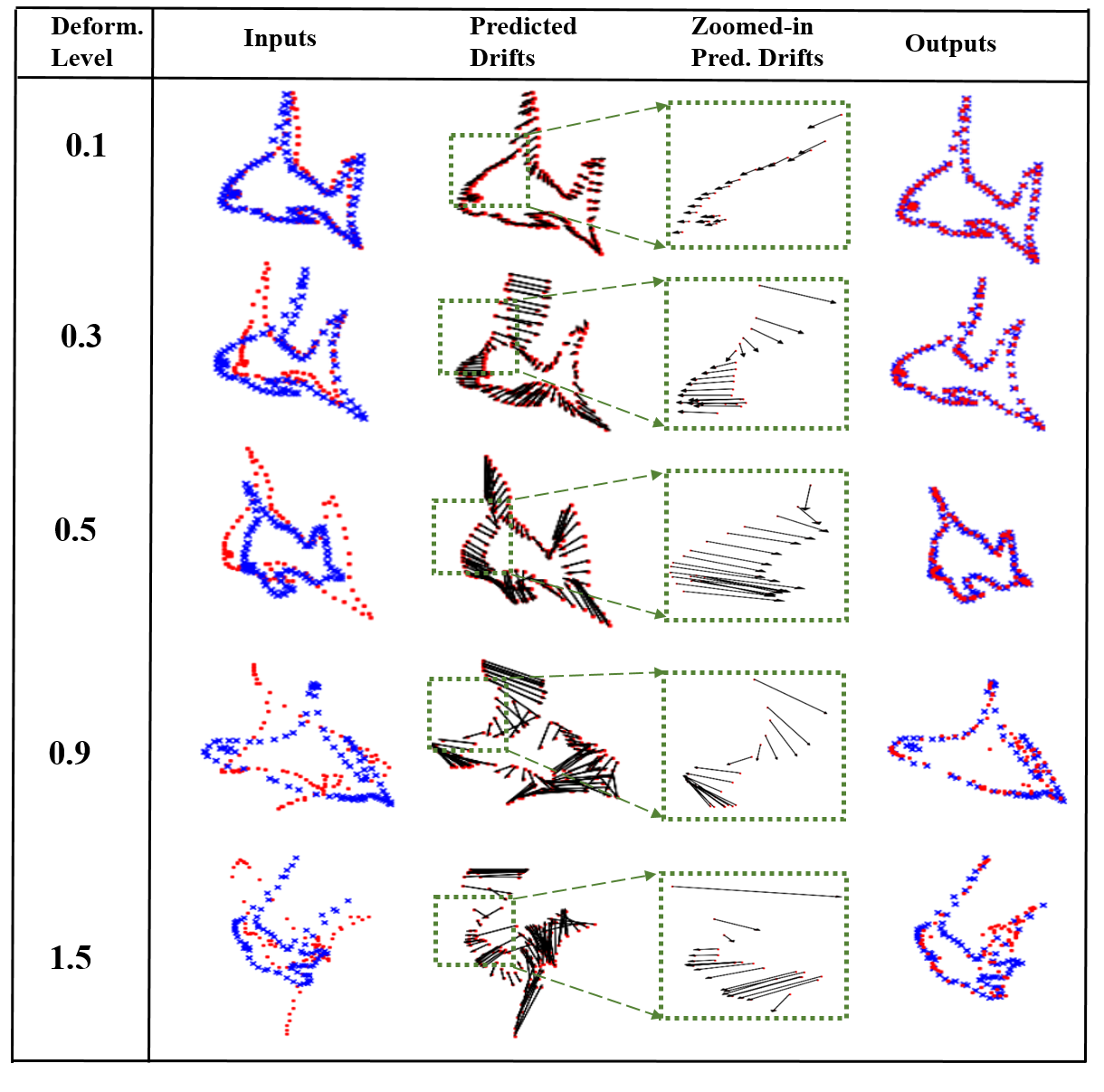}
\caption{The qualitative registration result for Fish shape at different deformation level. The target point sets are shown in blue while the source point sets are shown in red. The black lines are predicted coherent drifts for the source point set. Please zoom-in for better visualization.}
\label{f1}
\end{figure*}

A variety of different 2D and 3D shapes (i.e examples shown in Figure \ref{f1}, Figure \ref{other}, and Figure \ref{3d}) are used in the experiments to train and test the CPD-Net. In experiments, for synthesized dataset, we prepare the dataset as follows:
\begin{itemize}
    \item To prepare the deformable shape dataset (as shown in the first column row of Figure \ref{f1}), we simulate non-rigid geometric transformation on the normalized raw point sets by thin plate spline (TPS) \cite{bookstein1989principal} transformation with different deformation levels. The deformation level is defined as the perturbing degree of controlling points in TPS. Specifically, given the deformation level set at $l$ (e.g. 0.5), a Gaussian random shift with zero-mean and $2l$ standard deviation is generated to perturb the controlling points.\\
    
    \item To prepare the Gaussian Displacement (G.D.) noise dataset (as shown in the first row of Figure \ref{fig311}), we simulate the random displacement superimposed on a deformed point set (deformation level at $0.5$) by applying an increasing intensity of zero-mean Gaussian noise. The G.D. noise level is defined using the standard deviation of Gaussian. \\
    
    \item To prepare the Point Outlier (P.O.) noise for the shape (as shown in the second row of Figure \ref{fig311}), we simulate the outliers on the deformed point set (deformation level at $0.5$) by adding an increasing number of Gaussian outliers. The P.O. noise level is defined as a ratio of Gaussian outliers and the entire target point set. \\
    
    \item To prepare the Data Incompleteness (D.I.) noise (as shown in the second row of Figure \ref{fig311}), we remove an increasing number of neighboring points from target point set (deformation level at $0.5$). The D.I. noise level is defined as the percentage of the points removed from the entire target point set. \\
    
\end{itemize}


\begin{figure*}
\centering
\includegraphics[width=14cm,height=12.5cm]{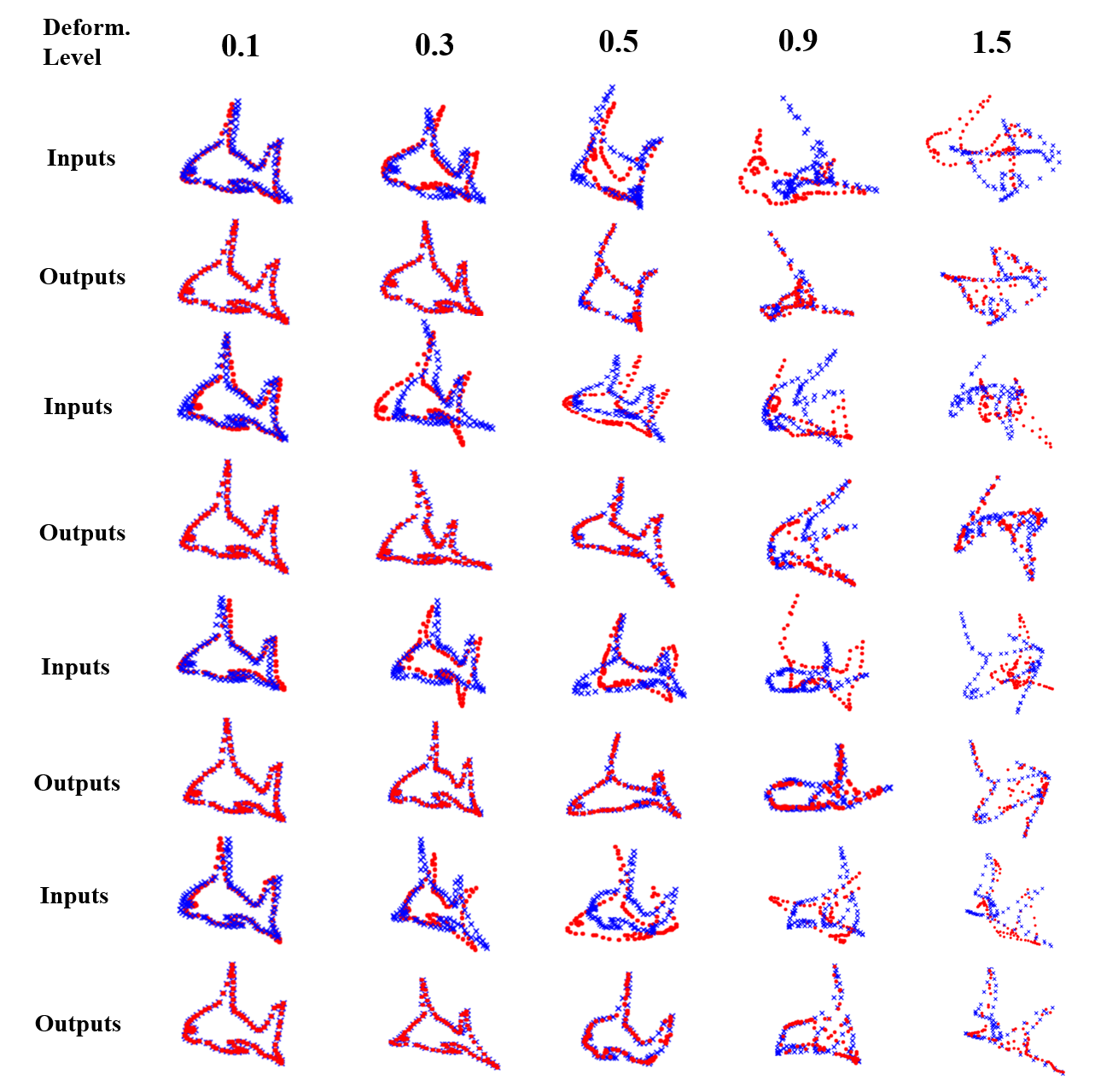}
\caption{The qualitative registration results for Fish shape at different deformation levels. The target point sets are shown in blue while the source point sets are shown in red. The black lines are predicted coherent drifts for the source point set. Please zoom-in for better visualization.}
\label{ff1}
\end{figure*}

\begin{figure*}
\centering
\includegraphics[width=14.3cm,height=4.2cm]{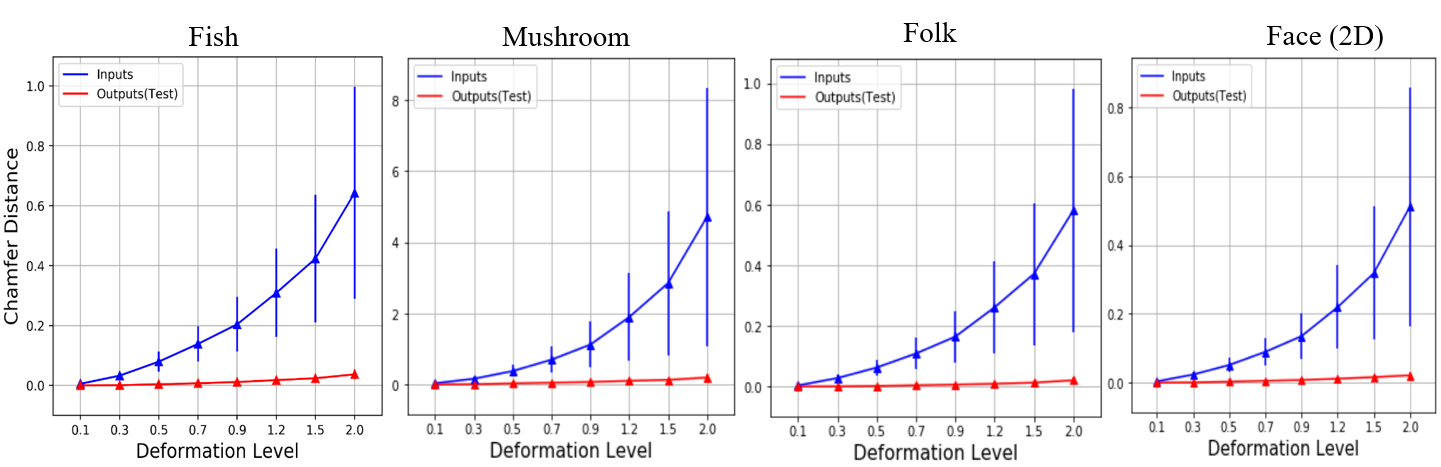}
\caption{The C.D. between source and target point sets, pre (blue line) and post (red line) registration.}
\label{oc}
\end{figure*}
\begin{table*}
\small
\begin{center}
\begin{tabular}{ccccccc}
\hline
Def. level&0.3&0.5&0.7&1.2&1.5&2.0 \\
 \hline
Fish&0.0008$\pm$0.0004&0.0037$\pm$0.0009&0.0072$\pm$0.0022&0.0178$\pm$0.0069&0.0239$\pm$0.0096&0.037$\pm$0.016\\
Mushroom &0.0006$\pm$0.0003&0.0031$\pm$0.0009&0.0051$\pm$0.0018&0.0103$\pm$0.0045&0.0129$\pm$0.0058&0.0196$\pm$0.0094\\
Fork &0.0002$\pm$0.0001&0.0014$\pm$0.0011&0.0038$\pm$0.0019&0.0089$\pm$0.0048&0.0126$\pm$0.0074&0.0203$\pm$0.0127 \\
Face (2D) &0.0005$\pm$0.0003&0.0028$\pm$0.0011&0.0053$\pm$0.0017&0.0114$\pm$0.0049&0.0158$\pm$0.0074&0.0213$\pm$0.01\\
\hline
\end{tabular}
\end{center}
\caption{Quantitative testing performance for 2D point set registration.}
\label{t1}
\end{table*}

\subsection{2D non-rigid point set registration}
In this experimental section, we demonstrate the point set registration performance of the CPD-Net on various categories of 2D shapes at different deformation levels.\\

\noindent{\textbf{Experiment Setting:}} In this experiment, we use four different types of 2D shapes (i.e. fork, face.) to prepare the dataset. For each shape, we first synthesize a set of $20k$ deformed shapes at each deformation level. The deformation level ranges from $0.3$ to $2.0$. To prepare the training dataset, for each type of shape at each deformation level, we split the synthesized dataset into two groups. We randomly choose a pair of shapes from group one to form $20k$ pairs of training. Similarly, we randomly choose two shapes from the other synthesized dataset to form $10k$ testing pairs. Note that there is no intersection between training and testing datasets. To evaluate the registration performance, we use the Chamfer Distance (C.D.) between the transformed source point set and target one as quantitative assessment, and we visualize the pairwise point sets before and after registration for qualitative assessment. We conduct two tests based on the shapes we used in the test. For test one, we use the commonly adopted fish shape. In test two, we use other 2D shape categories, including mushroom, face, and fork, to assess CPD-Net's performance, particularly on the none contour-based shapes.\\

\begin{figure}
\centering
\includegraphics[width=9.5cm,height=5.5cm]{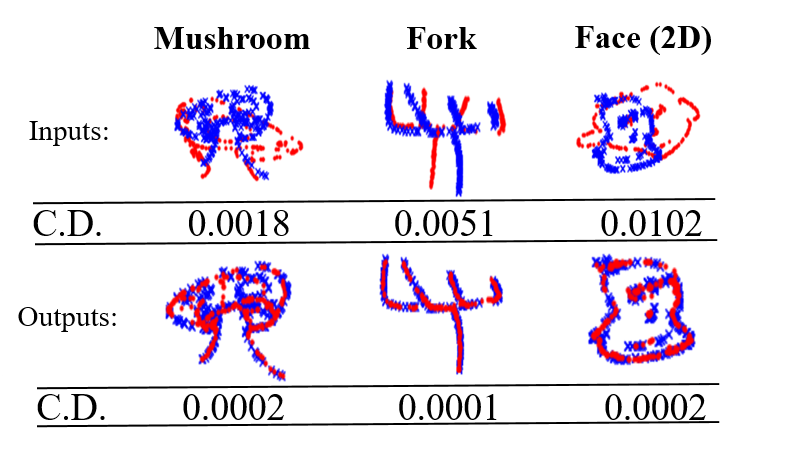}
\caption{Registration examples for Mushroom, Fork and Face shapes. The target point sets are shown in blue while the source point sets are shown in red. The corresponding C.D. score is listed underneath the registered point sets. }
\label{other}
\end{figure}

\noindent{\textbf{Results of Test 1:}}
After training CPD-Net, we applied the trained model to testing dataset prepared as described above. In Figure \ref{f1}, the first column illustrates the pair of source point set and target one before registration, where the target point sets are shown in blue while the source point sets are shown in red. The second column illustrates the predicted drift vectors (depicted by the black arrow) by our trained CPD-Net for each point in source point set. The third column is the zoom-in view of predicted drift vectors in a focused region. The fourth column shows the registered pairs of transformed source set and target one after registration. Shapes from different rows have different deformation levels from $0.1$ to $1.5$. From the fourth column, we can observe that CPD-Net can predict nearly perfect registration when the deformation level is smaller than $0.9$. While we increase the deformation to the level greater than $1.5$, the source and target point sets have significant shape structural variation, which dramatically increases the difficulty of point set registration. However, CPD-Net can still reliably transform the source point set to align the main portion of the shape of the target point set. In addition, it is interesting to observe from the second and third columns that source point set moves coherently as a whole towards the target one. This observation verifies that CPD-Net is able to predict a continuous smooth displacement field without the necessity to impose additional coherence constraint term. We further provide the mean and standard deviation of C.D. calculated from the $10K$ testing pairs at each deformation level for quantitative evaluation. In Figure \ref{oc}, we plot the mean and standard deviation for the set of C.D. between the source point set and the target one at all deformation levels. We can see from the comparison that the red curve (post registration) is consistently below the blue one (pre registration), which indicates CPD-Net is able to robustly register the source and target point set with a small C.D. Moreover, the red curve stays nearly flat as the deformation increases from $0.1$ to $2.0$, which indicates that CPD-Net's robust performance at high deformation level. The detailed qualitative result from randomly sampled pairs is demonstrated in Figure \ref{ff1} and the corresponding quantitative result is presented in Table \ref{t1}.\\

\noindent{\textbf{Results of Test 2:}} In this test, we further use CPD-Net to perform the non-rigid registration on other three types of 2D shapes including Mushroom, Fork and Face shapes as shown in Figure \ref{other}. To visualize the registration result, we compare the pair of testing shapes before and after registration at deformation level $0.5$ as shown in Figure \ref{other}. All randomly selected samples show nearly perfect registration. Similar to test 1, we present quantitative evaluation using C.D. metric for the non-rigid registration of those three types of shapes in Table \ref{t1}. Each row contains the mean and standard deviation of C.D. measurement for all testing pairs of shapes at the deformation level from $0.3$ to $2.0$. The first column of Figure \ref{other} illustrates the registration result on mushroom shape; the second column shows the result on fork and the third row demonstrates the registration result on 2D face shapes. We can observe that CPD-Net can predict the nearly perfect non-rigid transformation for all those three categories of shapes at a certain deformation level ($l=0.5$ here). For instance, the second row lists the mean and standard deviation of the set of Chamfer distance, which is a distance measurement between the transformed source point set (hand) and the target one. Based on the quantitative results shown in Figure \ref{oc}, for all the four shapes, CPD-Net demonstrates the remarkable performance of non-rigid registration as evidenced by the fact that the C.D. is dramatically reduced and consistently stays low after alignment at all deformation levels, especially after the deformation level increases to $1.5$ when the shape structure has been dramatically deteriorated (as shown in the last row of Figure \ref{f1}).
\begin{table*}
\footnotesize
\addtolength{\tabcolsep}{-5pt}
\begin{center}
\begin{tabular}{ccccc}
\hline
Def. level&0.3&0.5&0.8&1.2\\
\hline
Face(Inputs)&0.0133$\pm$0.0044&0.0337$\pm$0.0129&0.0778$\pm$0.0305&0.1621$\pm$0.0683\\
Face(Outputs)&0.001$\pm$0.0001&0.0021$\pm$0.0004&0.0055$\pm$0.0013&0.0123$\pm$0.0029 \\
Cat(Inputs)&0.0119$\pm$0.0032&0.0311$\pm$0.0083&0.0708$\pm$0.021&0.1471$\pm$0.0537\\
Cat(Outputs)&0.0009$\pm$0.0001&0.0024$\pm$0.0004&0.007$\pm$0.0013&0.013$\pm$0.003\\
\hline
\end{tabular}
\end{center}
\caption{Quantitative testing performance for 2D shapes.}
\label{tab441}
\end{table*}

\begin{figure*}
\centering
\includegraphics[width=12.4cm,height=7.5cm]{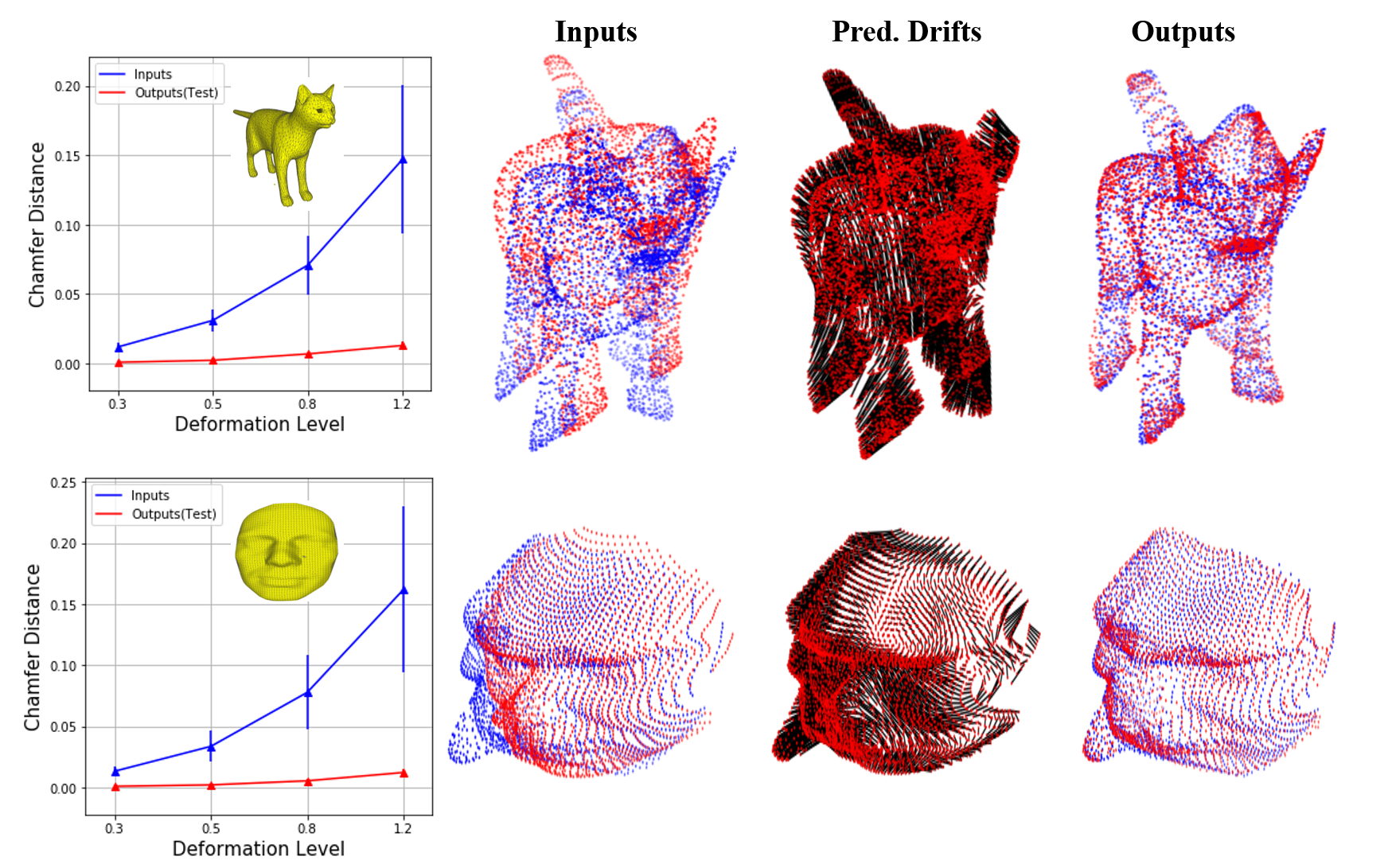}
\caption{The charts show C.D. between source and target point sets. Pre and post registration results are shown in blue lines and red lines respectively. Selected qualitative registration results are demonstrated in the right. The target point sets are shown in blue while the source point sets are shown in red. The black lines represent the predicted drifts for the source point set. Please zoom-in for better visualization.}
\label{3d}
\end{figure*}

\subsection{3D non-rigid point set registration }
In this experiment, we take a further step to investigate how well the CPD-Net performs 3D point set registration at different deformation levels since the 3D data have been gaining great attention in community with recent advancements in 3D acquisition and computation resources.\\

\noindent \textbf{Experimental Setting:} In this experiment, we use two categories of 3D shapes (i.e. 3D Face and 3D Cat) to prepare the dataset. Similar to 2D data preparation, we synthesize $10k$ training pairs of 3D shapes and $2k$ testing pairs for both 3D Face and Cat shapes at various deformation levels ($0.3$, $0.5$, $0.8$, and $1.2$). We use the same measurement methods for both qualitative and quantitative results (as shown in Figure \ref{3d}).

\begin{figure*}
\centering
\includegraphics[width=14.5cm,height=11.5cm]{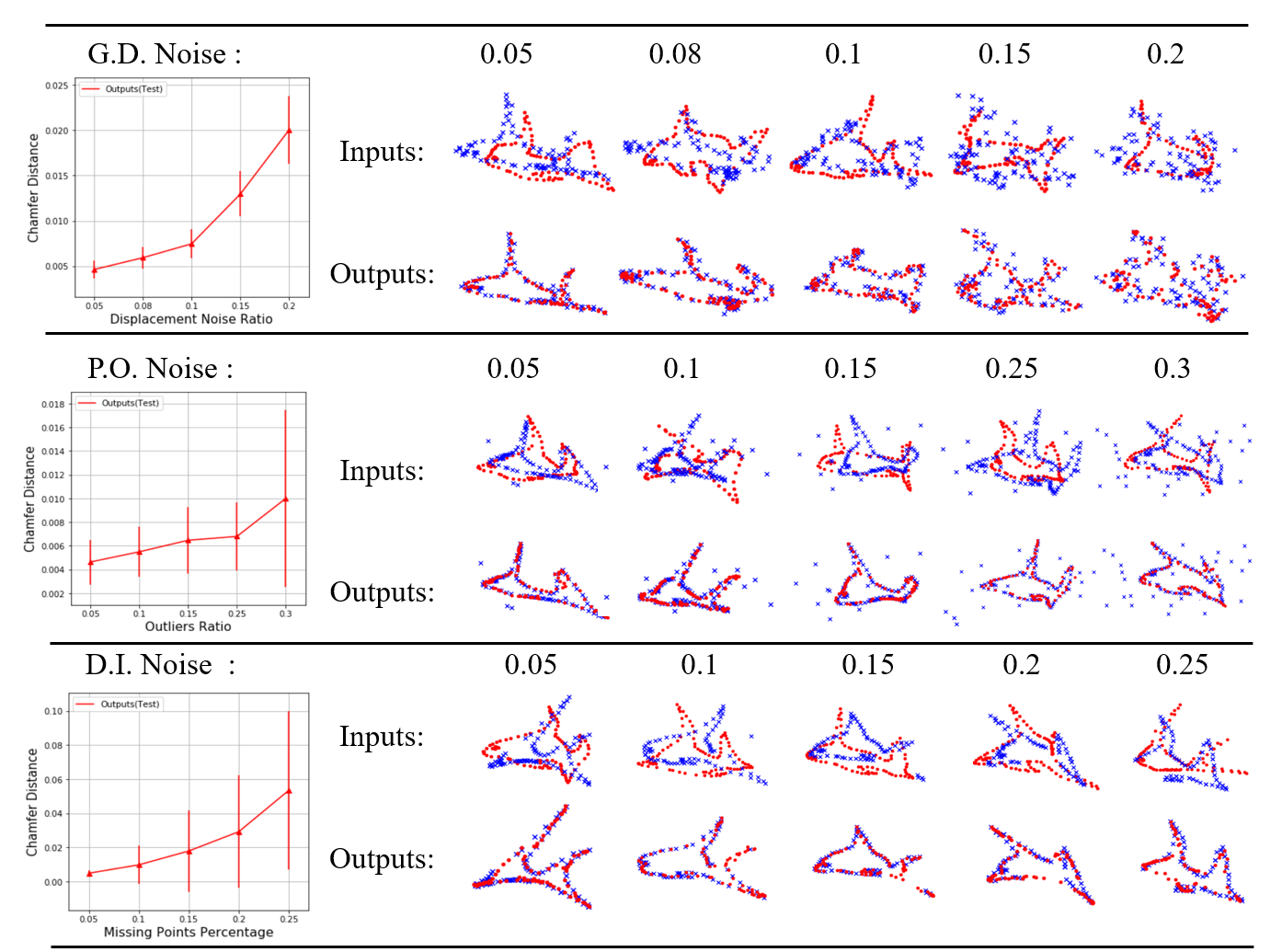}
\caption{The charts of C.D. between transformed source point set and target one in the presence of different level of G.D. noise, P.O. noise and D.I.noise are shown in left. The selected qualitative results are demonstrated in right. The target point sets are shown in blue while the source point sets are shown in red. Please zoom-in for better visualization.}
\label{fig311}
\end{figure*}
\noindent \textbf{Result:} In Figure \ref{3d}, we illustrate the quantitative evaluation curves on the left and visualize one registration cases at deformation level $0.3$ on the right. For both 3D cat and face shapes, CPD-Net demonstrates impressive performance with a quite small pairwise C.D. after registration, and the C.D. measurement remains consistently low while we increase the deformation level. At the level of $1.2$, the mean of C.D. between source and target point sets is nearly $10$ times less after alignment, which indicates that the trained CPD-Net is able to align most portion of the shapes between the source and target point sets. To further visualize the registration performance, we show the registration results on the right side of Figure \ref{3d}. In Figure \ref{3d}(right), the first column depicts the shape pair prior to registration, the second column depicts the coherent point drift vector for each point on the source point set, the third column depicts the shape pair after registration.
This figure clearly demonstrates that CPD-Net is able to predict 1) an accurate non-rigid transformation and 2) estimate a smoothly continuous 3D deformation field based on the coherent pint drift vectors shown on the second column (i.e. all of points move coherently as a whole towards the desired target positions by preserving both local and global continuity). The registration performance is particularly impressive to find the accurate non-rigid registration after the deformation level is greater than $l=0.8$, when the 3D structure of the shape objects are significantly deteriorated to a degree that human find it hard to register the pair of 3D objects.

\subsection{Resistance to Noise}
The data acquired by 3D sensors such as LIDAR sensor and laser scanner often suffer from various types of noises. An effective non-rigid registration method should be robust to those noise in addition to the structural variations as discussed in the previous section. Therefore, in this section, we focus on testing how well CPD-Net can predict the non-rigid registration from the noisy dataset. \\

\noindent \textbf{Experimental Setting:} In this section, we use fish shape data at a deformation level of $0.5$ to prepare the experimental noisy dataset. We simulate three types of noise (i.e. Gaussian Displacement (G.D.) noise, Point Outlier (P.O.) noise and Data Incompleteness (D.I.) noise). For each type, we gradually increase the level of noise added to the deformed target point set of fish dataset as shown in Figure \ref{fig311}). We prepare $10k$ pairs of source and target point sets for each type of noise for testing. As in the previous section, the same quantitative and qualitative performance measurement is used in this experiment.\\ 

\noindent \textbf{Result:} 
In this section, we illustrate the experimental result using the C.D. as the quantitative metric, and plot one pair of source and target point sets at different noise levels, pre and post registration. All experimental results are listed in Figure \ref{fig311}. As shown in Figure \ref{fig311}, we plot the quantitative evaluation curves on the left and visualize five registration cases at different noise level on the right. For the clean data, the mean of C.D. for the fish shape at a deformation level of $0.5$ is around $0.08$. We need to validate if CPD-Net can significantly reduce the C.D. for pairs of source and target point sets of noisy dataset after registration, and if CPD-Net can consistently keep the C.D. comparatively low when the noise level increases. 

\uline{For the G.D. noise}, in Figure \ref{fig311} the first row depicts the registration by our CPD-Net for the G.D. noise corrupted data. As we notice the plot, the C.D. after registration remains constantly lower than $0.08$, even when the G.D. noise increases to the level of $0.2$. CPD-Net can still predict the non-rigid transformation to align the source set (the red points) to the target one (the blue points) with a relatively small C.D. between two point sets, even though the shape was dramatically altered by the Gaussian Displacement noise. The shape was dramatically altered by the Gaussian noise which makes it difficult to recognize the overall shape as a fish. However, our CPD-Net can still predict the non-rigid registration to align the source set (the red points) to the target one (the blue points) with a relatively small C.D. between two point sets. 
\uline{For the P.O. noise}, as shown in the second row, outlier points are increasingly added to the target point set (blue ones) from left to right in a row. Different from the P.O. noise, we would like to check if CPD-Net is able to successfully ignore those outlier points during the registration process. The registration result is impressive that the main body of the source and target shapes can robustly be aligned to each other with a small C.D. between them after registration, especially when the P.O. noise level reaches as high as $0.3$. \uline{For D.I. noise}, as shown in the third row, an increasing number of points is removed from the target point set (blue ones). We would like to check if CPD-Net is able to successfully align the source point set to the incomplete portion of the target one. The visualization of pairwise registration result in the third row clearly demonstrates that CPD-Net is able to robustly align the source point set (red) to the incomplete target point set (blue). When D.I. noise level reaches $0.25$, the missing part is aligned with a straight line, which is less desired. But the aligned portions from transformed source point set and target one show consistent common geometric structure, which is not affected by the missing portion of the target point set. 


\begin{table}
\begin{center}
\small
\begin{tabular}{ccc}
\hline
Methods&C.D.&Time \\
 \hline
CPD \cite{myronenko2007non} (Train) &0.0039$\pm$0.0032&$\sim$ 22 \text{ hours}\\
Ours (Train) &0.0035$\pm$0.0008&$\sim$ 25 \text{ minutes}\\
\hline
CPD \cite{myronenko2007non} (Test) &0.0039$\pm$0.0033&$\sim$ 22 \text{ hours}\\
Ours (Test) &0.0037$\pm$0.0009&$\sim$ 15 \text{ seconds}\\
\hline
\end{tabular}
\end{center}
\caption{Performance and Time comparison with CPD.}
\label{t2}
\end{table}


\begin{table}
\begin{center}
\small
\begin{tabular}{ccc}
\hline
Methods&C.D.&Time \\
 \hline
CPD \cite{myronenko2007non} & $0.0042\pm0.0024$ &$\sim$ 4 \text{ hours}\\
Ours & $0.0024\pm0.0019$ & $\sim$ 40 \text{ seconds}\\
\hline
\end{tabular}
\end{center}
\caption{Performance and Time comparison with CPD on ShapeNet chair dataset.}
\label{tab4}
\end{table}

\begin{figure*}
\centering
\includegraphics[width=12.5cm]{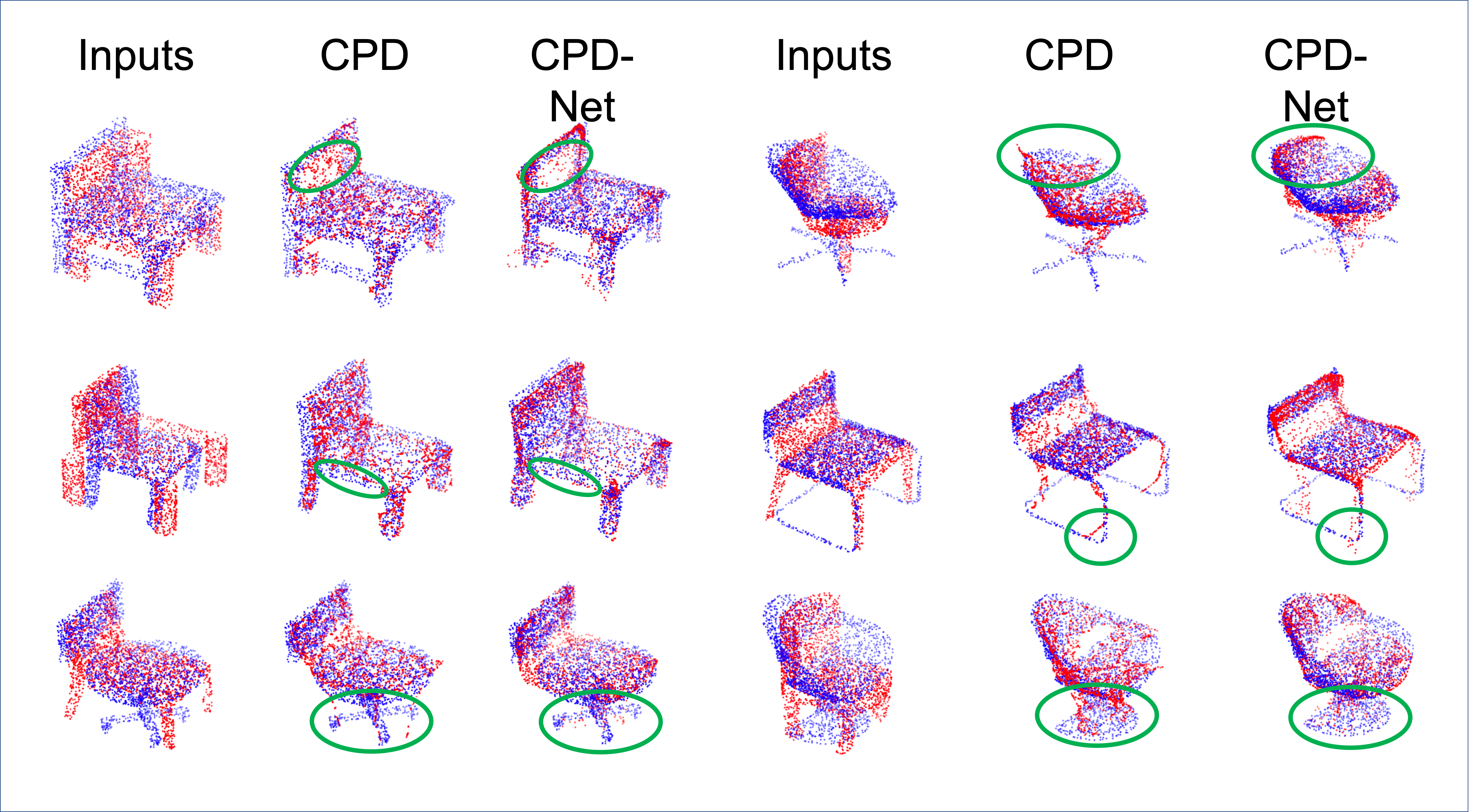}
\caption{Selected qualitative registration results on ShapeNet chair dataset. The target point sets are shown in blue while the source point sets are shown in red.}
\label{fig313}
\end{figure*}

\subsection{Comparison to CPD}
Different from previous efforts, the proposed CPD-Net is a learning-based non-rigid point set registration method, which can learn the registration pattern to directly predict the non-parametric geometric transformation for the point sets alignment. As a learning-based approach to predict the non-rigid registration, it is not applicable to have a direct comparison between CPD-Net and other existing non-rigid iterative registration methods. To compare our method to non-learning based iterative method (i.e. Coherent Point Drift (CPD) \cite{myronenko2007non}), we design the experiment as follows to assess both time and accuracy performance.\\

\noindent \textbf{Experimental Setting:} In this experiment, we use the fish shape at a deformation level of $0.5$ to prepare the dataset. We synthesize $20k$ pairwise source and target point sets to form the training set, and another $20k$ pairs to form the testing set. The CPD-Net is firstly trained with the $20k$ training dataset. The trained CPD-Net is then applied to directly predict registration for the $20k$ testing dataset. In contrast, CPD is directly applied to both $20k$ training and testing dataset. \\

\noindent \textbf{Result:} The C.D. based quantitative comparison is presented in the Table \ref{t2}. The first and the third row list the experimental results for CPD on training and testing dataset respectively. The second row lists the results for the CPD-Net on the training dataset, and the fourth row lists the results for the trained CPD-Net on the testing dataset. Based on the comparison between first and third rows, we can see that our model can achieve better performance (i.e. smaller C.D.) than that of CPD within a shorter time ($25$ minutes versus $22$ hours) to align $20k$ pairs. Unlike the CPD needs to start over a new iterative optimization process to register a new pair of shapes independently, CPD-Net actively learns the registration pattern from training and consequently become capable of handling real-time point set registration or a large volume dataset by direct predicting the geometric transformation. As shown in second and fourth row of the Table \ref{t1}, we notice that the trained CPD-Net is able to achieve nearly the same performance on the testing dataset as the training dataset, which indicates that CPD-Net has great generalization capability. The trained CPD-Net is able to achieve better performance than CPD on the same dataset with orders of magnitude less time ($15$ seconds versus $22$ hours).


\subsection{Registration of 3D real-world shapes}
In this section, we examine the performance of our proposed CPD-Net model for non-rigid registration of real-world 3D shapes. In comparison to the synthesized dataset, the real-world dataset is more challenging and the existence of the perfect desired transformation is not guaranteed. Since we propose the first learning-based non-rigid registration method, we do not have a fair baseline for comparison but we provide the results from CPD, a non-learning based iterative method. \\

\noindent \textbf{Experimental Setting:} We conducted our experiments on the ShapeNet chair dataset with 3,758 shapes. We randomly choose 2000 pairs from the training split for network training and 1000 pairs from the test split for network evaluation. One should note that there is no overlapping between training and testing dataset and the ground truth transformation is unknown and not used for the training process. The comparing method CPD was directly applied to the testing dataset. \\

\noindent \textbf{Result:} The qualitative results are demonstrated in Figure \ref{fig313}. As shown in Figure \ref{fig313}, our CPD-Net successfully transformed the source shapes to the target shape with small displacement. While the CPD model fails to get satisfying results on these cases. One should note that the results of our CPD-Net were directly generated in one network forward pass without iterative approximation used in CPD method. Table \ref{tab4} listed the quantitative results of both methods. As can be seen in this Table, our CPD-Net model achieved significantly better performance than CPD method with orders of magnitude less time(40 seconds versus 4 hours).

\section{Conclusion}
In this paper, we report our latest research effort towards non-rigid point set registration. The proposed method, Coherent Point Drift Network (CPD-Net), is presented to the research community as a new alternative solution for non-rigid point registration. Existing approaches often iteratively search for the optimal geometric transformation to register a given pair of point sets and are commonly driven by minimizing a pre-defined alignment loss function. In contrast, the proposed CPD-Net can actively learn the registration pattern from training dataset as non-parametric geometric transformation in an unsupervised manner. Consequently, CPD-Net can predict the desired geometric transformation to align a pair of unseen point sets. To the best of our knowledge, this is the first effort in the community that an algorithm can actually learn to register point sets from training in an unsupervised fashion. The CPD-Net can be extended to other modality data such as 2D and 3D image and a high-dimensional multimedia data. We will present the extension of the CPD-Net in separate works.

\begin{rezabib}
@article{ma2016non,
  title={Non-rigid point set registration by preserving global and local structures},
  author={Ma, Jiayi and Zhao, Ji and Yuille, Alan L},
  journal={IEEE Transactions on image Processing},
  volume={25},
  number={1},
  pages={53--64},
  year={2016},
  publisher={IEEE}
}

@article{jian2011robust,
  title={Robust point set registration using gaussian mixture models},
  author={Jian, Bing and Vemuri, Baba C},
  journal={IEEE transactions on pattern analysis and machine intelligence},
  volume={33},
  number={8},
  pages={1633--1645},
  year={2011},
  publisher={IEEE}
}

@article{bai2007skeleton,
  title={Skeleton pruning by contour partitioning with discrete curve evolution},
  author={Bai, Xiang and Latecki, Longin Jan and Liu, Wen-Yu},
  journal={IEEE transactions on pattern analysis and machine intelligence},
  volume={29},
  number={3},
  year={2007},
  publisher={IEEE}
}

@article{bai2008path,
  title={Path similarity skeleton graph matching},
  author={Bai, Xiang and Latecki, Longin Jan},
  journal={IEEE transactions on pattern analysis and machine intelligence},
  volume={30},
  number={7},
  pages={1282--1292},
  year={2008},
  publisher={IEEE}
}

@article{myronenko2009image,
  title={Image registration by minimization of residual complexity},
  author={Myronenko, Andriy and Song, Xubo},
  year={2009},
  publisher={IEEE}
}

@inproceedings{wu2012online,
  title={Online robust image alignment via iterative convex optimization},
  author={Wu, Yi and Shen, Bin and Ling, Haibin},
  booktitle={2012 IEEE Conference on Computer Vision and Pattern Recognition},
  pages={1808--1814},
  year={2012},
  organization={IEEE}
}

@article{ma2014robust,
  title={Robust point matching via vector field consensus.},
  author={Ma, Jiayi and Zhao, Ji and Tian, Jinwen and Yuille, Alan L and Tu, Zhuowen},
  journal={IEEE Trans. image processing},
  volume={23},
  number={4},
  pages={1706--1721},
  year={2014}
}
@inproceedings{ling2005deformation,
  title={Deformation invariant image matching},
  author={Ling, Haibin and Jacobs, David W},
  booktitle={Computer Vision, 2005. ICCV 2005. Tenth IEEE International Conference on},
  volume={2},
  pages={1466--1473},
  year={2005},
  organization={IEEE}
}
@inproceedings{klaus2006segment,
  title={Segment-based stereo matching using belief propagation and a self-adapting dissimilarity measure},
  author={Klaus, Andreas and Sormann, Mario and Karner, Konrad},
  booktitle={Pattern Recognition, 2006. ICPR 2006. 18th International Conference on},
  volume={3},
  pages={15--18},
  year={2006},
  organization={IEEE}
}

@article{maintz1998survey,
  title={A survey of medical image registration},
  author={Maintz, JB Antoine and Viergever, Max A},
  journal={Medical image analysis},
  volume={2},
  number={1},
  pages={1--36},
  year={1998},
  publisher={Elsevier}
}

@inproceedings{besl1992method,
  title={Method for registration of 3-D shapes},
  author={Besl, Paul J and McKay, Neil D},
  booktitle={Sensor Fusion IV: Control Paradigms and Data Structures},
  volume={1611},
  pages={586--607},
  year={1992},
  organization={International Society for Optics and Photonics}
}
@inproceedings{raguram2008comparative,
  title={A comparative analysis of RANSAC techniques leading to adaptive real-time random sample consensus},
  author={Raguram, Rahul and Frahm, Jan-Michael and Pollefeys, Marc},
  booktitle={European Conference on Computer Vision},
  pages={500--513},
  year={2008},
  organization={Springer}
}
@article{yuille1988computational,
  title={A computational theory for the perception of coherent visual motion},
  author={Yuille, Alan L and Grzywacz, Norberto M},
  journal={Nature},
  volume={333},
  number={6168},
  pages={71},
  year={1988},
  publisher={Nature Publishing Group}
}
@book{sonka2014image,
  title={Image processing, analysis, and machine vision},
  author={Sonka, Milan and Hlavac, Vaclav and Boyle, Roger},
  year={2014},
  publisher={Cengage Learning}
}

@article{tam2013registration,
  title={Registration of 3D point clouds and meshes: a survey from rigid to nonrigid.},
  author={Tam, Gary KL and Cheng, Zhi-Quan and Lai, Yu-Kun and Langbein, Frank C and Liu, Yonghuai and Marshall, David and Martin, Ralph R and Sun, Xian-Fang and Rosin, Paul L},
  journal={IEEE transactions on visualization and computer graphics},
  volume={19},
  number={7},
  pages={1199--1217},
  year={2013},
  publisher={Institute of Electrical and Electronics Engineers, Inc., 3 Park Avenue, 17 th Fl New York NY 10016-5997 United States}
}

@inproceedings{chui2000new,
  title={A new algorithm for non-rigid point matching},
  author={Chui, Haili and Rangarajan, Anand},
  booktitle={Computer Vision and Pattern Recognition, 2000. Proceedings. IEEE Conference on},
  volume={2},
  pages={44--51},
  year={2000},
  organization={IEEE}
}

@article{bookstein1989principal,
  title={Principal warps: Thin-plate splines and the decomposition of deformations},
  author={Bookstein, Fred L.},
  journal={IEEE Transactions on pattern analysis and machine intelligence},
  volume={11},
  number={6},
  pages={567--585},
  year={1989},
  publisher={IEEE}
}

@article{ma2015robust,
  title={Robust L2E Estimation of Transformation for Non-Rigid Registration.},
  author={Ma, Jiayi and Qiu, Weichao and Zhao, Ji and Ma, Yong and Yuille, Alan L and Tu, Zhuowen},
  journal={IEEE Trans. Signal Processing},
  volume={63},
  number={5},
  pages={1115--1129},
  year={2015}
}

@inproceedings{wang2016path,
  title={Path following with adaptive path estimation for graph matching},
  author={Wang, Tao and Ling, Haibin},
  booktitle={Thirtieth AAAI Conference on Artificial Intelligence},
  year={2016}
}

@inproceedings{zhou2015multi,
  title={Multi-image matching via fast alternating minimization},
  author={Zhou, Xiaowei and Zhu, Menglong and Daniilidis, Kostas},
  booktitle={Proceedings of the IEEE International Conference on Computer Vision},
  pages={4032--4040},
  year={2015}
}

@inproceedings{su2015multi,
  title={Multi-view convolutional neural networks for 3d shape recognition},
  author={Su, Hang and Maji, Subhransu and Kalogerakis, Evangelos and Learned-Miller, Erik},
  booktitle={Proceedings of the IEEE international conference on computer vision},
  pages={945--953},
  year={2015}
}

@inproceedings{sharma2016vconv,
  title={Vconv-dae: Deep volumetric shape learning without object labels},
  author={Sharma, Abhishek and Grau, Oliver and Fritz, Mario},
  booktitle={Computer Vision--ECCV 2016 Workshops},
  pages={236--250},
  year={2016},
  organization={Springer}
}
@inproceedings{maturana2015voxnet,
  title={Voxnet: A 3d convolutional neural network for real-time object recognition},
  author={Maturana, Daniel and Scherer, Sebastian},
  booktitle={Intelligent Robots and Systems (IROS), 2015 IEEE/RSJ International Conference on},
  pages={922--928},
  year={2015},
  organization={IEEE}
}

@inproceedings{verma2018feastnet,
  title={FeaStNet: Feature-Steered Graph Convolutions for 3D Shape Analysis},
  author={Verma, Nitika and Boyer, Edmond and Verbeek, Jakob},
  booktitle={CVPR 2018-IEEE Conference on Computer Vision \& Pattern Recognition},
  year={2018}
}

@inproceedings{masci2015geodesic,
  title={Geodesic convolutional neural networks on riemannian manifolds},
  author={Masci, Jonathan and Boscaini, Davide and Bronstein, Michael and Vandergheynst, Pierre},
  booktitle={Proceedings of the IEEE international conference on computer vision workshops},
  pages={37--45},
  year={2015}
}

@article{qi2017pointnet,
  title={Pointnet: Deep learning on point sets for 3d classification and segmentation},
  author={Qi, Charles R and Su, Hao and Mo, Kaichun and Guibas, Leonidas J},
  journal={Proc. Computer Vision and Pattern Recognition (CVPR), IEEE},
  volume={1},
  number={2},
  pages={4},
  year={2017}}

@inproceedings{zeng20173dmatch,
  title={3dmatch: Learning local geometric descriptors from rgb-d reconstructions},
  author={Zeng, Andy and Song, Shuran and Nie{\ss}ner, Matthias and Fisher, Matthew and Xiao, Jianxiong and Funkhouser, Thomas},
  booktitle={Computer Vision and Pattern Recognition (CVPR), 2017 IEEE Conference on},
  pages={199--208},
  year={2017},
  organization={IEEE}
}

@inproceedings{myronenko2007non,
  title={Non-rigid point set registration: Coherent point drift},
  author={Myronenko, Andriy and Song, Xubo and Carreira-Perpin{\'a}n, Miguel A},
  booktitle={Advances in Neural Information Processing Systems},
  pages={1009--1016},
  year={2007}
}

@inproceedings{rocco2017convolutional,
  title={Convolutional neural network architecture for geometric matching},
  author={Rocco, Ignacio and Arandjelovic, Relja and Sivic, Josef},
  booktitle={Proc. CVPR},
  volume={2},
  year={2017}
}
@inproceedings{ma2013robust,
  title={Robust estimation of nonrigid transformation for point set registration},
  author={Ma, Jiayi and Zhao, Ji and Tian, Jinwen and Tu, Zhuowen and Yuille, Alan L},
  booktitle={Proceedings of the IEEE Conference on Computer Vision and Pattern Recognition},
  pages={2147--2154},
  year={2013}
}

@article{zhang2016understanding,
  title={Understanding deep learning requires rethinking generalization},
  author={Zhang, Chiyuan and Bengio, Samy and Hardt, Moritz and Recht, Benjamin and Vinyals, Oriol},
  journal={arXiv preprint arXiv:1611.03530},
  year={2016}
}

@article{girosi1995regularization,
  title={Regularization theory and neural networks architectures},
  author={Girosi, Federico and Jones, Michael and Poggio, Tomaso},
  journal={Neural computation},
  volume={7},
  number={2},
  pages={219--269},
  year={1995},
  publisher={MIT Press}
}

@article{yuille1989mathematical,
  title={A mathematical analysis of the motion coherence theory},
  author={Yuille, Alan L and Grzywacz, Norberto M},
  journal={International Journal of Computer Vision},
  volume={3},
  number={2},
  pages={155--175},
  year={1989},
  publisher={Springer}
}

@inproceedings{nair2010rectified,
  title={Rectified linear units improve restricted boltzmann machines},
  author={Nair, Vinod and Hinton, Geoffrey E},
  booktitle={Proceedings of the 27th international conference on machine learning (ICML-10)},
  pages={807--814},
  year={2010}
}

@inproceedings{bai2016gift,
  title={Gift: A real-time and scalable 3d shape search engine},
  author={Bai, Song and Bai, Xiang and Zhou, Zhichao and Zhang, Zhaoxiang and Jan Latecki, Longin},
  booktitle={Proceedings of the IEEE Conference on Computer Vision and Pattern Recognition},
  pages={5023--5032},
  year={2016}
}

@inproceedings{balakrishnan2018unsupervised,
  title={An Unsupervised Learning Model for Deformable Medical Image Registration},
  author={Balakrishnan, Guha and Zhao, Amy and Sabuncu, Mert R and Guttag, John and Dalca, Adrian V},
  booktitle={Proceedings of the IEEE Conference on Computer Vision and Pattern Recognition},
  pages={9252--9260},
  year={2018}
}

@article{xu2015empirical,
  title={Empirical evaluation of rectified activations in convolutional network},
  author={Xu, Bing and Wang, Naiyan and Chen, Tianqi and Li, Mu},
  journal={arXiv preprint arXiv:1505.00853},
  year={2015}
}

@article{fan2016point,
  title={A point set generation network for 3d object reconstruction from a single image},
  author={Fan, Haoqiang and Su, Hao and Guibas, Leonidas},
  journal={arXiv preprint arXiv:1612.00603},
  year={2016}
}
@inproceedings{fan2017point,
  title={A Point Set Generation Network for 3D Object Reconstruction from a Single Image.},
  author={Fan, Haoqiang and Su, Hao and Guibas, Leonidas J},
  booktitle={CVPR},
  volume={2},
  number={4},
  pages={6},
  year={2017}
}

@inproceedings{ioffe2015batch,
  title={Batch normalization: Accelerating deep network training by reducing internal covariate shift},
  author={Ioffe, Sergey and Szegedy, Christian},
  booktitle={International Conference on Machine Learning},
  pages={448--456},
  year={2015}
}
\end{rezabib}

\bibliographystyle{IEEEtran}

\end{document}